\documentclass[aps,pre,twocolumn,amsfonts,amssymb,amsmath]{revtex4-1}
\usepackage{graphicx,xcolor,soul}

\newcommand{\dd}{\mathrm{d}}

\newcommand{\IInt}[3]{\int_{#2}^{#3}\dd #1\;}

\renewcommand{\vec}[1]{\mathbf #1}
\newcommand{\mat}[1]{\mathbf #1}

\newcommand{\al}{\alpha}
\newcommand{\lam}{\lambda}
\newcommand{\vhi}{\varphi}
\newcommand{\om}{\omega}
\newcommand{\tW}{\tilde{\mat W}}
\newcommand{\ps}{p^\text{s}}
\newcommand{\pe}{p^\text{eq}}
\newcommand{\pc}{p^\text{cg}}
\newcommand{\id}{\mathbf 1}

\begin{document}

\title{Non-equilibrium Markov state modeling of periodically driven biomolecules}

\author{Fabian Knoch}
\author{Thomas Speck}
\affiliation{Institut f\"ur Physik, Johannes Gutenberg-Universit\"at Mainz,
  Staudingerweg 7-9, 55128 Mainz, Germany}

\begin{abstract}
  Molecular dynamics simulations allow to study the structure and dynamics of single biomolecules in microscopic detail. However, many processes occur on time scales beyond the reach of fully atomistic simulations and require coarse-grained multiscale models. While systematic approaches to construct such models have become available, these typically rely on microscopic dynamics that obey detailed balance. In vivo, however, biomolecules are constantly driven away from equilibrium in order to perform specific functions and thus break detailed balance. Here we introduce a method to construct Markov state models for systems that are driven through periodically changing one (or several) external parameter. We illustrate the method for alanine dipeptide, a widely used benchmark molecule for computational methods, exposed to a time-dependent electric field.
\end{abstract}

\maketitle


\section{Introduction}

The accurate modeling of large-scale complex systems such as proteins is a central challenge of computational (bio)chemistry. Molecular dynamics (MD) simulations have become an invaluable tool to gain microscopic insights into these pathways, and to calculate free energies and rate constant~\cite{frenkel}. However, fully atomistic unbiased simulations are still limited to hundreds of nanoseconds, while processes such as protein folding occur on much longer time scales. Markov state models (MSMs) are able to bridge time scales from nanoseconds up to milliseconds and beyond, and have been successfully applied to investigate folding pathways of large proteins~\cite{bowman2010,voelz2010}, but also ligand-binding kinetics~\cite{plattner2015}. The dynamics of MSMs obeys a discrete-time master equation on a discrete space of long-lived (metastable) mesostates. MSMs are constructed from an ensemble of MD trajectories, but can also be augmented with experimental data~\cite{rudz16}. A key ingredient is the detailed balance condition linking forward and backward transition probabilities to energy changes, which drastically facilitates the construction of MSMs~\cite{prinz2011}. Recently, some efforts have been undertaken to extend MSMs towards driven systems such as modelling sliding friction~\cite{pelle16,teruzzi17}, time-dependent external fields~\cite{wang2015}, and polymer collapse in shear flow~\cite{knoch2017}.

In living cells, large proteins also perform specific functions such as assembling the ribosome and the generation of forces, \emph{e.g.} the motor protein kinesin~\cite{kolo07}. These molecular machines operate cyclically through converting chemical energy (typically through hydrolyzing adenosine triphosphate), which drives conformational changes. For example, in F$_1$-ATPase these changes lead to the rotation of a central stalk through which the motor can exert mechanical forces and thus perform work on the environment~\cite{noji97}. The typical approach is to model such motors as operating in a non-equilibrium steady state (NESS) with stochastic dynamics that break detailed balance~\cite{liep07,zimm12}. Much progress has been made in formulating a coherent thermodynamic framework including the ever-present thermal fluctuations, which goes by the name stochastic thermodynamics~\cite{seif12}. In contrast, traditional designs of (macroscopic) engines are based on periodic variations of, \emph{e.g.} the temperature, with the system returning to thermal equilibrium once the external stimulus is turned off. Stochastic thermodynamics has been extended to such periodic processes~\cite{bran15,ray17}. In the following, we will exploit that periodic driving can be represented as a NESS~\cite{raz16,rots17}, which greatly facilitates the theoretical analysis.

Here we develop a method to systematically construct periodically driven coarse-grained MSMs from atomistic data. To this end, we combine previous work on non-equilibrium Markov state modeling (NE-MSM) for time-dependent protocols~\cite{knoch18} and non-equilibrium steady states~\cite{knoch2015,knoch2017}. Specifically, we construct a series of equilibrium fine-grained MSMs at different values of the external parameter. Exploiting the mapping to a NESS, coarse-graining is then performed in the space of cycles, thereby respecting the microscopic pathways along which entropy is produced. In Sec.~\ref{sec:theory} we provide some theoretical background on the mapping and stochastic thermodynamics~\cite{seif12}. In Sec.~\ref{sec:impl}, we describe the details of the implementation before discussing results in Sec.~\ref{sec:results} for a specific illustration of our method: alanine dipeptide in a time-dependent, periodic electric field. A possible application of this method is the efficient calculation of infrared spectra of biomolecules~\cite{krimm86}.


\section{Theory}
\label{sec:theory}

\subsection{Mapping to NESS}

We are concerned with classical systems that are well described by stochastic transitions between a (possibly very large) set of discrete \emph{microstates}. These states represent small volumes in phase space. The time evolution is assumed to be governed by the master equation
\begin{equation}
  \label{eq:master}
  \dot p_j(t) = \sum_i p_i(t)W_{ij}(t),
\end{equation}
where $p_i(t)$ is the probability to occupy microstate $i$, and $W_{ij}(t)$ with $i\neq j$ is the transition rate to jump from state $i$ to state $j$. The diagonal entries $W_{ii}=-\sum_{j\neq i}W_{ij}$ ensure probability conservation. Eq.~\eqref{eq:master} is solved formally by the fundamental matrix $\mat K(t)$, which, in matrix notation, obeys
\begin{equation}
  \label{eq:fundamental}
  \dot{\mat K}(t) = \mat K(t)\cdot\mat W(t)
\end{equation}
with initial condition $\mat K(0)=\id$ and $(\mat W)_{ij}=W_{ij}$. For a time-dependent rate matrix $\mat W(t)$, no general closed solution of Eq.~\eqref{eq:fundamental} exists.

Here we are interested in rates with a periodic time dependence, $\mat W(t+T)=\mat W(t)$, where $T$ is the period. After a transient period, the actual probabilities acquire the same periodicity as the driving, $\vec p(t+T)=\vec p(t)$. We seek a stationary Markov processes described by a time-independent rate matrix $\tW$ that mimics this (quasi)-stationary periodic process. The problem is typically formulated as follows: Given the time-averaged occupation probabilities and fluxes
\begin{gather}
  \label{eq:avg}
  \bar p_i \equiv \frac{1}{T}\IInt{s}{t}{t+T} p_i(s), \\
  \bar\Phi_{ij} \equiv \frac{1}{T}\IInt{s}{t}{t+T} p_i(s)W_{ij}(s),
\end{gather}
respectively, what are the rates $\tilde W_{ij}$ that yield the same probabilities $\bar p_i$ and currents $\bar J_{ij}\equiv\bar\Phi_{ij}-\bar\Phi_{ji}$? As realized by Zia and Schmittmann~\cite{zia07}, while the currents fix the antisymmetric part, there is some freedom in choosing the symmetric part of the fluxes. Two choices have been discussed recently, one that preserves the entropy production along edges~\cite{raz16} and one that preserves the fluctuations of the currents~\cite{rots17}.

Here we follow a different route and make use of \emph{Floquet's theorem}~\cite{kuchment2012}, which states that the solution of Eq.~\eqref{eq:fundamental} can be split into a periodic and a nonperiodic part leading to
\begin{equation}
  \label{eq:floquet}
  \mat K(t+nT) = \mat K(t)\cdot[\mat K(T)]^n
\end{equation}
with integer $n$. Clearly, $\mat P=\mat K(T)$ is the transition matrix of a time-discrete Markov chain. We assume that the Markov chain can be reproduced by a time-continuous Markov process with \emph{time-independent} rate matrix $\tW$ so that $\mat P\equiv\exp(\tW T)$. By construction, both $\mat P$ and $\tW$ have the same eigenvectors and thus the same stationary probabilities $\vec\ps$. It is easy to check that
\begin{equation}
  \bar{\vec p}\cdot\mat P = \frac{1}{T}\IInt{s}{t}{t+T} \vec p(s+T)
  = \bar{\vec p}
\end{equation}
independent of $t$ with $\vec p(t)\cdot\mat P=\vec p(t+T)$. Hence, we conclude that $\vec\ps=\bar{\vec p}$ due to the uniqueness of the stationary solution of $\mat P$. However, the fluxes
\begin{equation}
  \label{eq:phi}
  \Phi_{ij} \equiv \ps_i\tilde W_{ij},
\end{equation}
and thus the currents $J_{ij}\equiv\Phi_{ij}-\Phi_{ji}$, are not necessarily the same as the time-averaged fluxes and currents of the periodic process. The computational advantage of our approach is that we do not have to measure currents in the periodic process, but it is sufficient to estimate the transition probabilities $P_{ij}$, a task that is well suited for Markov state modeling.

\subsection{Determining the effective dynamics}
\label{sec:eff}

The challenge is to determine the effective transition probabilities $\mat P$ in non-trivial systems with many microstates. One approach described by Wang and Sch\"utte~\cite{wang2015} is to explicitly incorporate some time-dependent, periodic signal into molecular dynamics simulations and to build a non-equilibrium Markov chain from these simulations. This fine-grained model is described by the transition matrix $\mat P$ and is then further coarse-grained into metastable states identified from milestoning~\cite{schutte11,sarich2014}. However, this approach is somewhat limited because changing the protocol, or even the period, requires to rerun the atomistic MD simulations.

\begin{figure}[b!]
  \centering
  \includegraphics[scale=0.8]{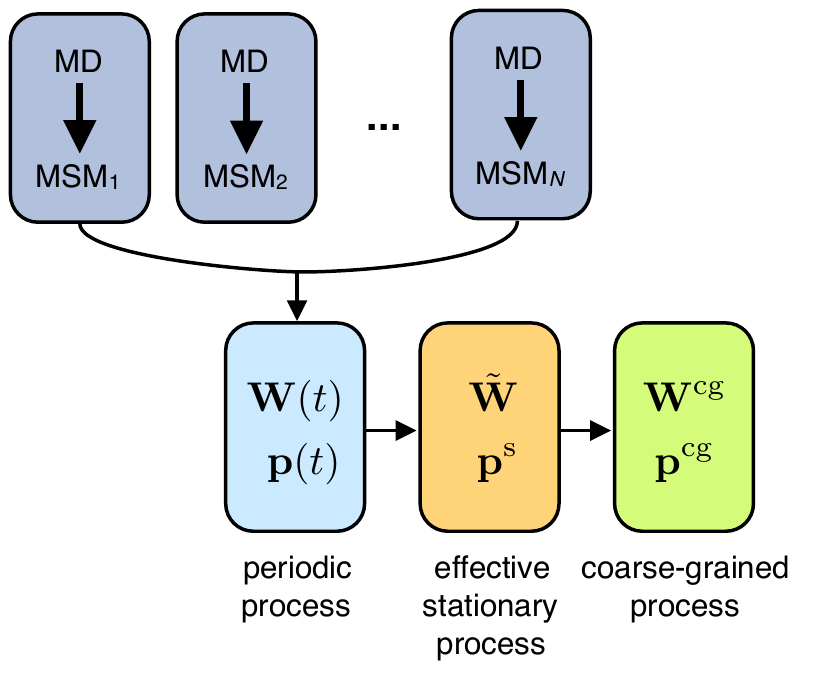}
  \caption{\textbf{Schematic work flow.} A series of fine-grained MSMs is constructed for different values of the external field. These are used to approximate the time-dependent rate matrix $\mat W(t)$, which is mapped onto an effective stationary process. The last step is the construction of the coarse-grained model with a reduced set of states.}
  \label{fig:scheme}
\end{figure}

Our strategy is different. From stationary atomistic MD simulations, and using standard tools, we construct a series of $N$ \emph{equilibrium} Markov state models parametrized by the value of the external parameter $\lam$~\cite{knoch18}. The procedure is sketched in Fig.~\ref{fig:scheme}(a). For each of these MSMs indexed by $k=1\dots N$ with $\lam_k$, we determine the rate matrix $\mat W^{(k)}$. The time-dependent matrix $\mat W(t)$ is then approximated step-wise by these matrices through discretizing the protocol $\lam(t)\approx\lam_k$ for $t_k<t<t_{k+1}$. This allows us to implement different protocols $\lam(t)$ and, in particular, different periods $T$ reusing the same equilibrium MSMs. The transition probabilities
\begin{equation}
  \label{eq:solve_K}
  \mat P = \mat K(T) \approx \id + \sum_{k=0}^{N-1} \IInt{s}{t_k}{t_{k+1}}
  \mat K(s)\cdot\mat W^{(k)}(s)
\end{equation}
are obtained through integrating Eq.~(\ref{eq:fundamental}).

Formally, the rate matrix $\tW$ can be obtained by employing the matrix logarithm, $\tW=\frac{1}{T}\ln(\mat P)$. However, this expression is not guaranteed to return a correct rate matrix $\tW$, \emph{i.e.}, every non-diagonal element is non-negative, which is known as the embedding problem. It is not yet known what conditions $\mat P$ must fulfill to be embeddable. When, however, a given matrix $\mat P$ is not embeddable, it is still useful to determine an auxiliary rate matrix approximating the time-continuous dynamics. One approach described in Ref.~\cite{israel2001} is to approximate the matrix logarithm by its series expansion
\begin{align}
  \label{eq:convert}
  \tW &= \frac{1}{T}\ln(\mat P) =\frac{1}{\tau}\log{(\id + \mat P - \id)}\\
  \nonumber
  &\approx \frac{1}{T}\left[(\mat P - \id) - \frac{(\mat P - \id )^2}{2} + \frac{(\mat P - \id)^3}{3} - ... \right].
\end{align}
The expansion can be computed recursively, while we stop either if any non-diagonal entry of the auxiliary matrix becomes negative, or the change in the next order approximation is small enough. Note that the linear approximation always returns a correct rate matrix. More powerful methods to calculate matrix logarithms are available and will be explored in the future.

\subsection{Detailed balance and entropy production}

The original dynamics obeys detailed balance at each point in time in the sense that for the vectors $\vec\pe$ obeying $\vec\pe\cdot\mat W(t)=0$ we have $\pe_iW_{ij}(t)=\pe_jW_{ji}(t)$ for any value of $t$ (which merely parametrizes the rates). Hence, freezing the protocol, the system would return to thermal equilibrium. For periodic driving, entropy is produced with rate $\dot S_\text{tot}$ due to the actual probabilities $\vec p(t)$ being different from $\vec\pe$.

Mathematically, the network of states $\{i\}$ can be represented as an undirected graph, where the edges correspond to possible transitions. A single system jumps randomly from state to state so that the ensemble average obeys the master equation~\eqref{eq:master}. In this graph we can identify \emph{cycles} $\al\equiv(i\to j\to \dots \to i)$, which are defined as ordered sets of states, at the end of which the starting state is reached again and all other states are visited exactly once. Stationary Markov processes that break detailed balance, such as the effective dynamics obtained in Sec.~\ref{sec:eff}, imply non-zero currents $J_{ij}$ which have to flow along these cycles (which is a consequence of Kirchhoff's current law $\sum_j J_{ij}=0$). These currents imply an entropy production with average rate~\cite{seif12}
\begin{equation}
  \label{eq:entropy}
  \dot S = \frac{1}{2}\sum_{ij} J_{ij}A_{ij} \geqslant 0,
\end{equation}
where we have introduced the microscopic affinities $A_{ij}\equiv\ln(\Phi_{ij}/\Phi_{ji})$. In contrast, in the original time-dependent process probability can be temporarily accumulated within a period, see Ref.~\cite{rots17} for a discussion. In the following, we ignore this extra contribution and focus on the entropy production $\dot S$ defined in Eq.~(\ref{eq:entropy}). In particular, our coarse-graining scheme will preserve this entropy production of the effective dynamics.

\subsection{Thermodynamically consistent coarse-graining}
\label{sec:cg}

As already mentioned, in the effective stationary process all entropy has to be produced in cycles. Coarse-graining, \emph{i.e.} the removal of states, most likely will also remove cycles and thus reduce the entropy production~\cite{pugl10}. We have devised a coarse-graining scheme that circumvents this problem through performing the coarse-graining in the space of cycles~\cite{knoch2015,knoch2017}.

Our main tool is the cycle-flux decomposition~\cite{kalpazidou2007,alta12a}. We decompose fluxes as
\begin{equation}
  \label{eq:decomp}
  \Phi_{ij} = \Phi^0_{ij} + J_{ij} = \Phi^0_{ij} + \sum_{\al\ni(i\to j)} \vhi_\al
\end{equation}
with cycle weights $\vhi_\al$, where the sum is over all cycles $\al$ that contain the oriented edge $i\to j$. Here, $\Phi^0_{ij}=\Phi^0_{ji}$ is the symmetric part of the fluxes. Details on the algorithm and the efficient determination of the cycle weights are provided in appendix~\ref{sec:decomp}. Inserting Eq.~(\ref{eq:decomp}) into Eq.~(\ref{eq:entropy}), we obtain $\dot S=\sum_\al\vhi_\al A_\al$ with cycle affinities $A_\al$ (summing the edge affinities $A_{ij}$ along cycle $\al$).

The second step is to identify similar cycles and group them into communities, each of which is represented by a single cycle (details on how to do this are given in Sec.~\ref{sec:cycle}). We keep only the set of these cycle representatives and delete all remaining cycles. The surviving states (being part of at least one cycle representative) then form the coarse-grained model. The new transition rates are obtained under two restrictions: (i)~the mean entropy production rate $\dot S$ is preserved and (ii)~all edge affinities $A_{ij}$ are preserved. The entropy production rate of the $l$th community follows as
\begin{equation}
  \label{eq:sdot_com}
  \dot S_l = \sum_{\al\in C_l} A_\al\vhi_\al,
\end{equation}
where $C_l$ contains all cycles belonging to the $l$th cycle community. The cycle affinity of the cycle representative (which is an element of $C_l$) is indicated as $A_l$. The only quantity that can be modified then is the cycle weight $\vhi_l\to\vhi^\text{cg}_l$ of the cycle representative, which is chosen as
\begin{equation}
  \dot S_l \stackrel{!}{=} A_l \vhi^\text{cg}_l
  \quad\Rightarrow\quad \vhi^\text{cg}_l = \frac{\dot S_l}{A_l}
\end{equation}
such that $\dot{S}_l$ is preserved. The symmetric contribution to the fluxes can be shown to be given by~\cite{knoch2015}
\begin{equation}
  \Phi^{0,\text{cg}}_{ij} = \frac{\sum_{l\ni(i\to j)}\vhi^\text{cg}_l}{e^{A_{ij}}-1}.
\end{equation}
Finally, preserving affinities $A_{ij}$ implies that the logarithm of the ratio of conjugated nonzero rates $\ln W^\text{cg}_{ij}/W^\text{cg}_{ji}$ is conserved, which is fulfilled if $\ps_i/\ps_j=\pc_i/\pc_j$. We rescale all probabilities by the same factor so that $\sum_i p_i^\text{cg}=1$. Having obtained $\Phi_{ij}^\text{cg}$ and $p_i^\text{cg}$, the final elements of the coarse-grained rate matrix are determined through
\begin{equation}
  W_{ij}^\text{cg} = \frac{\Phi_{ij}^\text{cg}}{p_i^\text{cg}}.
\end{equation}
Further details can be found in Ref.~\cite{knoch2015}.


\section{Implementation}
\label{sec:impl}

\subsection{Molecular dynamics simulations}

\begin{figure}[b!]
  \centering
  \includegraphics[width=\linewidth]{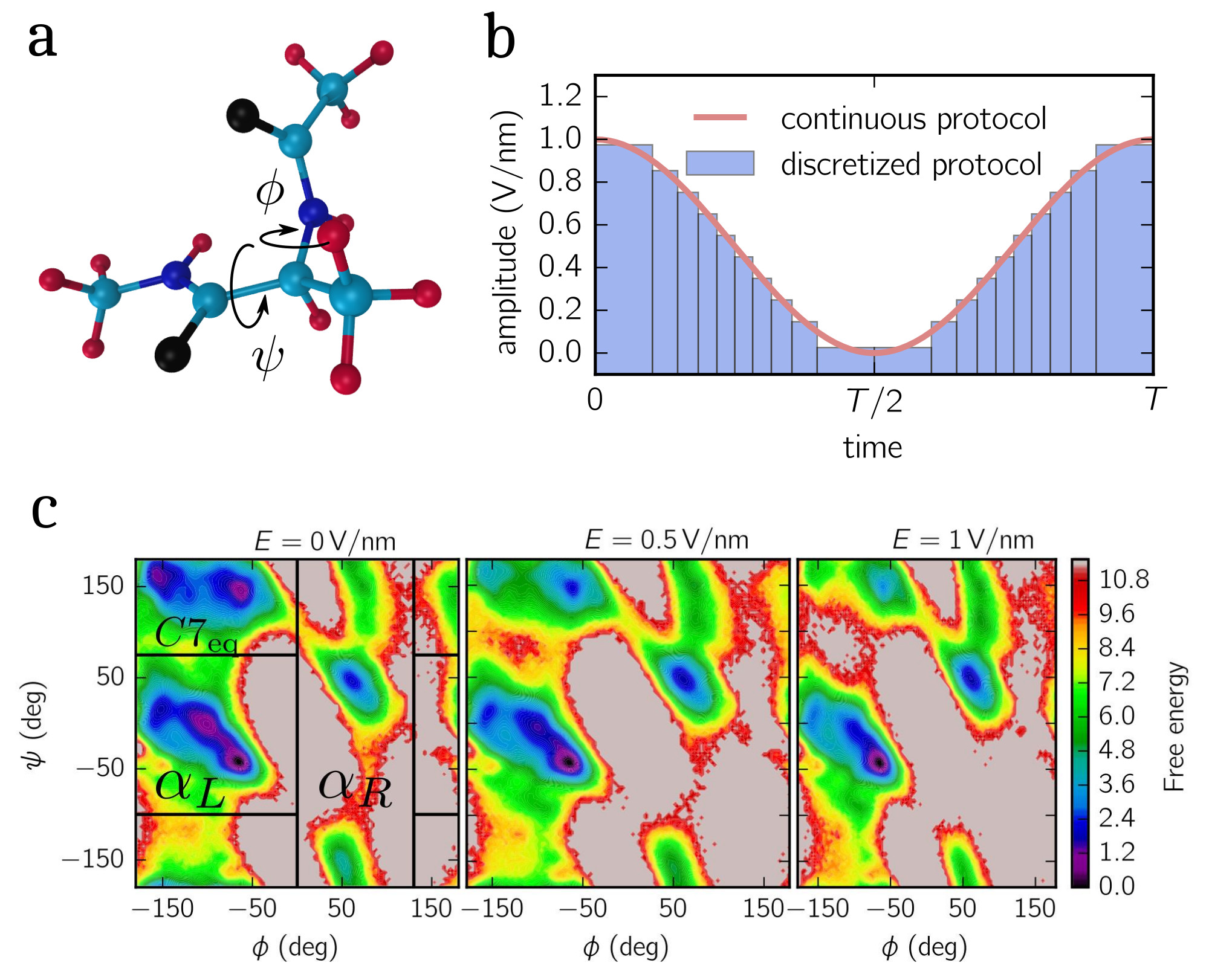}
  \caption{\textbf{Alanine dipeptide.} (a)~Exemplary snapshot of the atomistic structure indicating the dihedral angles $\phi$ and $\psi$. (b)~Continuous protocol $E(t) = \frac{1}{2}[\cos{(2\pi t/T)}+1]$ and its discretization. (c)~Ramachandran plot of free energies. From left to right: A constant electric field $E$ in positive $x$-direction is applied with magnitudes $1\,$V/nm, $0.5\,$V/nm, and $0\,$V/nm. The colors in all three plots follow the same scale indicated on the right.}
  \label{fig:alanine}
\end{figure}

As a concrete example, we study alanine dipeptide in aqueous solution, which serves as a paradigm for computational studies of free energy calculations and rate estimations of protein dynamics~\cite{esque2015,chodera2007,oliveia2007,vega2009}. The molecular conformations are well characterized by its two dihedral angles $\phi$ and $\psi$ illustrated in Fig.~\ref{fig:alanine}(a).

All molecular dynamics simulations of alanine dipeptide have been performed employing the molecular dynamics software Gromacs 5.1.2~\cite{gromacs5}. The alanine dipeptide molecule is modeled with the CHARMM27~\cite{charmm27} force field, dissolved in 362 TIP3P~\cite{jorgensen1983} water molecules and set up in a 2.25~nm~$\times$~2.25~nm$~\times~$2.25~nm box with periodic boundary conditions. Long range electrostatics were treated using particle mesh ewald summation with cubic interpolation and Fourier grid spacing of 0.16~nm. The cutoff for all short-ranged interactions was set to 1.0~nm. All hydrogen-involving covalent bonds were constraint by the LINCS algorithm~\cite{lincs}. For the time step we chose 2~fs. The temperature was set to 300~K using velocity-rescaling~\cite{bussi2007} thermostat with $\tau_T=1~$ps, while for the isotropic pressure coupling we used the Parrinello-Rahman~\cite{parrinello1981} barostat with $\tau_p=2~$ps.

\subsection{Constructing the equilibrium MSM}
\label{sec:eq}

MD simulations are performed at a constant electric field. We harvest 500 independent equilibrium trajectories ($\approx 10~$ns per trajectory) for each of the ten considered field strengths [see Fig.~\ref{fig:alanine}(b)] and record configurations every 1~ps. All simulations are started from the same initial configuration with the first 2~ns of each trajectory being discarded. From these trajectories, we construct a series of equilibrium MSMs based on both dihedral angles ($\psi,\phi$). Instead of directly discretizing the dihedral angles, we employ the cos/sin projection for both angles, returning a four-dimensional space, \emph{i.e.},
\begin{equation}
  \label{eq:dihedral_space}
  \begin{pmatrix}
    \phi\\\psi  
  \end{pmatrix}
  \quad\longmapsto\quad
  \begin{pmatrix}
    \cos\phi \\ \sin\phi \\ \cos\psi \\ \sin\psi
  \end{pmatrix}. 
\end{equation}
The benefit of doubling the dimensionality is that, in contrast to the dihedral space, the cos/sin space offers a distance metric which allows the application of distance-based clustering algorithms and is required for our definition of cycle centers and diameters.

To further improve the configuration space representation, we employ the time-lagged independent components analysis (TICA)~\cite{perez2013} with lag time $\tau_\text{tica}=1$~ps. TICA is an orthogonal linear transformation similar to a principal component analysis (PCA). However, its orthogonal components point in the direction in which the slowest conformational changes occur for a given input lag time $\tau_\text{tica}$. Although the kinetic variance ($\approx 96\%$) of the TICA transformation suggests a reduced two-dimensional configuration space, we keep all four TICA dimensions as it allows us to back-transform from TICA space into dihedral space. Note that our analysis can be equally applied for a reduced configuration space. However, for illustrative purposes it is instructive to keep all dimensions.

The transformed configuration space is then clustered (using the $k$-means algorithm) into 250 states represented by centroids, which are positions in the underlying configuration space (here TICA space). We have chosen 250 states as a compromise between a too fine-grained rate matrix (with possibly badly estimated entries due to insufficient statistic for states that are almost never visited) and a too-coarse rate matrix. Because the output of the $k$-means algorithm depends on the starting configuration of centroids (zeroth iteration), we perform the clustering with 50 different randomly selected configurations and select the discretization exhibiting the smallest error. We verified that the best three discretizations do not show any qualitative difference. The final centroids are labeled as $R^{(i)}_k$ indicating the $i$th component in TICA space of the $k$th centroid, with $k=1\dots250$ and $i=1\dots4$.

For constructing the equilibrium MSMs, we identify along every recorded trajectory a sequence of states, \emph{i.e.}, every position in TICA space is associated with its nearest (Euclidean distance) centroid. To estimate the rate matrix $\mat W^{(k)}$, we employ the maximum likelihood estimator~\cite{mcgib2015} provided in the software package MSMBuilder~\cite{msmbuilder}, for which use a lag time $\tau=2~$ps.

\subsection{Cycle space}
\label{sec:cycle}

To identify similarities among cycles, we need to quantify their properties. Here we follow the approach of Ref.~\cite{knoch2017} and focus on ``geometric properties'', but other measures are possible. Before defining these cycle space coordinates, it is convenient to introduce the indicator function
\begin{equation}
  \chi^i_\alpha = \begin{cases}
    1&\text{if state } i\text{ is in cycle }\alpha\\
    0&\text{otherwise}                
  \end{cases}
\end{equation}
whether or not a given state is part of a specific cycle.

We define two geometric collective variables to characterize cycles: the \emph{cycle centers} (or rather their $i$th component)
\begin{equation}
  \label{eq:cycle_com}
  c^{(i)}_\alpha \equiv \left(\frac{1}{|\alpha|}\sum_{k}R^{(i)}_k\chi^k_\alpha \right),
\end{equation}
where $|\alpha|=\sum_k \chi^k_\alpha $ denotes the length (number of states) of cycle $\alpha$ and $R^{(i)}_k$ the $i$th component of the $k$th centroid, and the \emph{cycle diameters}
\begin{equation}
  \label{eq:cycle_diameter}
  d^{(i)}_\alpha \equiv 
  \max_{k,l}\Bigg\{\chi^k_\alpha\chi^l_\alpha~\Big|R^{(i)}_k-R^{(i)}_l\Big|\Bigg\}.
\end{equation}
Together with the cycle affinities $A_\al$, the cycle centers and cycle diameters of all cycles form the cycle space. Note that, although strictly speaking the cycle centers and diameters are four-dimensional, we only make use of the first two dimensions in accordance with the two largest TICA components. Based on these features, we identify similiar cycles and group them into communities, each of which is represented by a single cycle representative. The details of the algorithm are given in Appendix~\ref{sec:repr}, which yields the final coarse-grained model with rates $\mat W^\text{cg}$.


\section{Results}
\label{sec:results}

\subsection{Static electric field}

Applying a static electric field leads to a new equilibrium state. To show the influence of a static electric field, we compute the free energy landscape projected onto the dihedral angles for three different field strengths shown in Fig.~\ref{fig:alanine}(c). Without an applied electric field, we identify three different regions known as the extended conformation of the backbone ($C7_\text{eq}$), left-handed ($\al_L$) and right-handed ($\al_R$) $\al$-helical conformers. Note that other studies split the $C7_\text{eq}$~\cite{wang2015} or $\alpha_R$~\cite{esque2015} regions, or both~\cite{chodera2007}, into further subregions. The classification employed here is the coarsest that still allows to identify the dominant conformations.

Already when applying a static electric field, a significant shift in the depths of the free energy basins can be observed. For $E=0~$V/nm the $C7_\text{eq}$ and $\alpha_L$ conformations are equally dominant, while the $\alpha_R$ conformations are comparatively rare. However, for $E=1~$V/nm the $\alpha_R$ conformations become more dominant while $C7_\text{eq}$ and $\alpha_L$ are less populated.

\subsection{Oscillating electric field}

The periodic protocol that drives the system out of equilibrium has the physical meaning of an oscillating electric field caused, for example, by a laser beam. For concreteness, we assume a polarized electric field $\vec E(t)=E(t)\vec e_x$ with periodic $E(t)=E_0[\cos(\om t)+1]$, which is characterized by the amplitude $E_0$ and frequency $\om=2\pi/T$ with $T$ the oscillation period~\footnote{We choose the electric field to be positive because the $(\phi,\psi$) configuration space would otherwise be degenerated with respect to positive/negative field directions, making it impossible to compare both approaches within the $(\phi,\psi)$ representation.}.

We start our modeling approach by discretizing the time-continuous electric field $E(t)$ into 20 steps. Both continuous and discrete electric fields are shown in Fig.~\ref{fig:alanine}(b). Because of the symmetry of the cosine function, however, the number of effective discretized field strengths are reduced by a factor of two. As described in Sec.~\ref{sec:eq}, we conduct independent MD simulations for these 10 different field strengths, each yielding an equilibrium MSM defined through the rate matrix $\mat W^{(k)}=\mat W(E_k)$.


\subsection{Stochastic pumping}

\begin{figure}[b!]
  \centering
  \includegraphics[width=\linewidth]{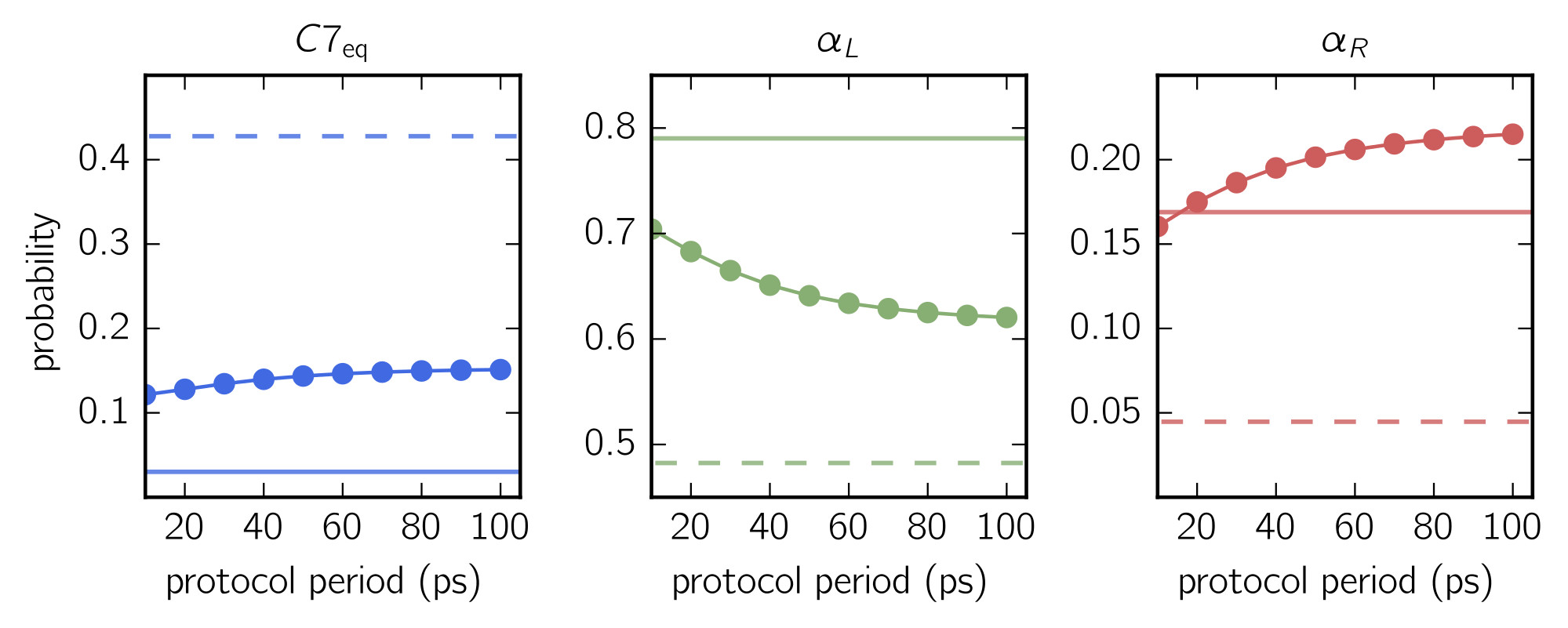}
  \caption{\textbf{Stochastic pumping.} The cumulated steady-state probabilities $P_X$ of the metastable sets ($C7_\text{eq}$, $\alpha_L$, $\alpha_R$) are shown as a function of driving period $T$. Vertical lines correspond to equilibrium probabilities for $E=1~$V/nm (solid) and $E=0~$V/nm (dashed).}
  \label{fig:prob_protocol}
\end{figure}

To study the effect of an oscillating field on the dynamics of alanine dipeptide, we solve Eq.~\eqref{eq:solve_K} for different periods and extract the corresponding effective rate matrix $\tW$ from which we compute the respective steady-state probability distributions. In Fig.~\ref{fig:prob_protocol} the cumulated stationary probability distribution for all three metastable sets ($C7_\text{eq}$, $\alpha_L$, $\alpha_R$) is shown for different driving periods $T$, whereas the solid/dashed lines illustrate the equilibrium distributions for no ($E = 0~$V/nm) and the strongest ($E=1~$V/nm) statically applied electric field. The cumulated probabilities are defined by summing over the corresponding steady-state probabilities,
\begin{equation}
  \label{eq:1}
  P_X \equiv \sum_{i\in X} \ps_i
\end{equation}
with $X=C7_\text{eq},\al_L,\al_R$. All three probabilities seem to saturate for longer periods. The probabilities $P_{\alpha_L}$ and $P_{\alpha_R}$ increase with longer periods $T$, while $P_{C7_\text{eq}}$ decreases. The absolute change, however, is moderate with about 10-15\%.

Interestingly, the stationary distributions compared with the equilibrium distributions do not follow any general trend, \emph{e.g.} $P_{\alpha_L}$ is located between both static distributions with $P_{\alpha_L}(E=0\,\text{V/nm})<P_{\alpha_L}<P_{\alpha_L}(E=1\,\text{V/nm})$. For $P_{C7_\text{eq}}$ we observe the opposite trend, \emph{i.e.}, $P_{C7_\text{eq}}(E=0\,\text{V/nm})>P_{C7_\text{eq}}>P_{C7_\text{eq}}(E=1\,\text{V/nm})$, and finally for $P_{\alpha_R}$ none is true as both values are exceeded. In particular, the latter is of importance as it demonstrates the concept of stochastic pumping~\cite{astumian2011}, where an oscillating protocol is used to increase the occupation of certain molecular conformations above their equilibrium values.

\begin{figure}[b!]
  \centering
  \includegraphics[width=0.7\linewidth]{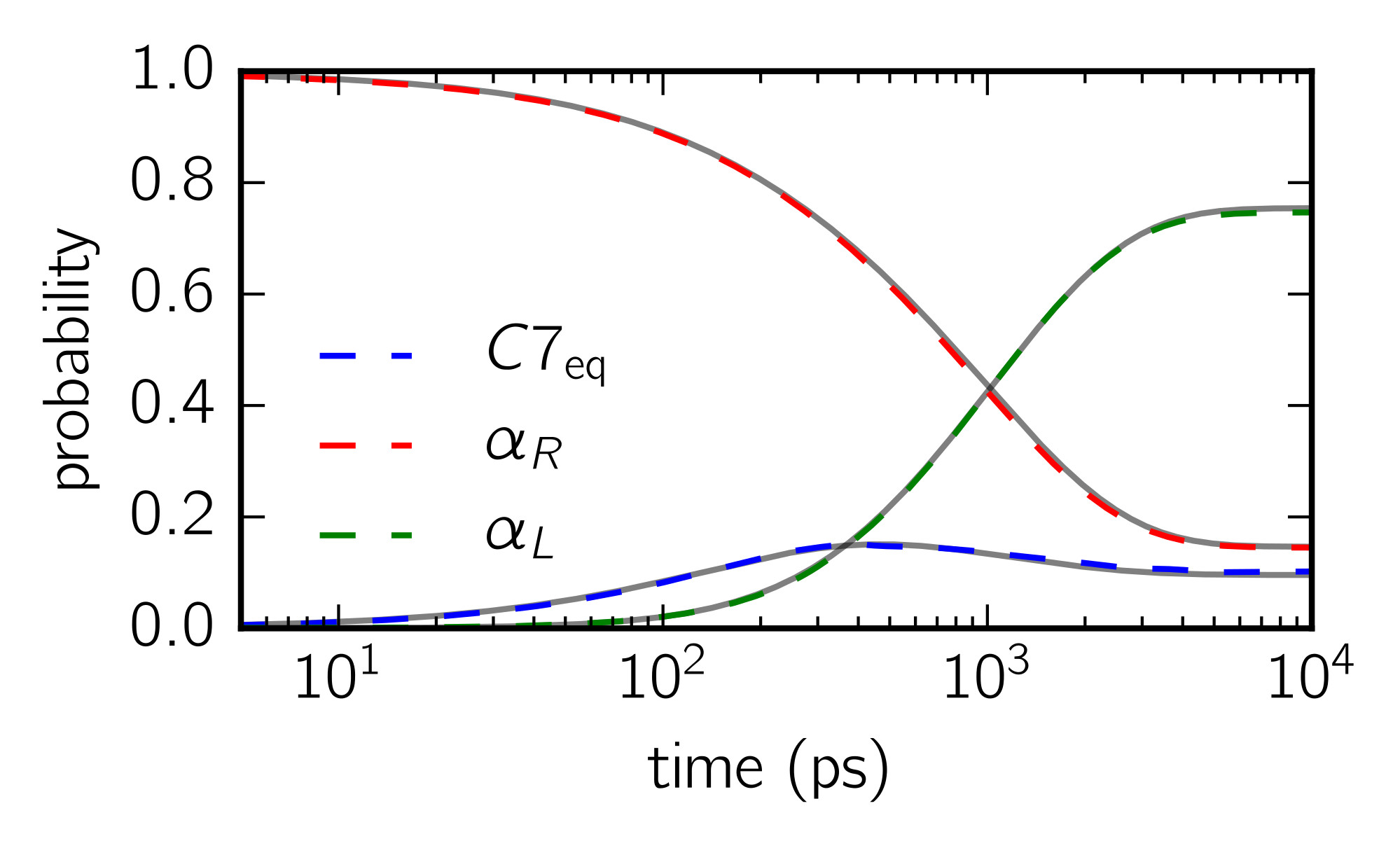}
  \caption{\textbf{Time-dependent probabilities.} Time evolution of the cumulated probability $P_X(t)$ for all three metastable sets (for the definition see main text) following the approaches developed in this work (colored dashed lines) and introduced in Ref.~\cite{wang2015} (gray lines). The oscillation period of the electric field is $T=5$~ps.}
  \label{fig:compare}
\end{figure}

To compare the approach introduced in this study with the complementary approach recently introduced by Wang and Sch\"utte~\cite{wang2015} (cf. Sec.~\ref{sec:eff}), we conduct 1000 MD simulations (10~ns each, removing the first 2~ns for equilibration) explicitly including the oscillating electric field with oscillation period $T=5~$ps. From the recorded trajectories, we directly estimate the matrix $\mat K(T)$ (lag time 5~ps) based on the same configuration space discretization as described in Sec.~\ref{sec:eq}, and convert it to $\tW$ following Eq.~\eqref{eq:convert}. To compare the quality of both approaches, we start in the initial state
\begin{equation}
  p_i(0) = \begin{cases}
    1/\sum_{j\in\alpha_R},& i\in \alpha_R\\
    0. & \text{otherwise}.          
  \end{cases}
\end{equation}
and monitor the time evolution of the cumulative probability distribution of all three metastable sets, which are now time-dependent. Fig.~\ref{fig:compare} depicts the time evolution of all three probability distributions for both approaches (gray lines represent the approach introduced by ref.~\cite{wang2015} and colored lines represent the approach introduced in this study), illustrating that both approaches are in excellent agreement. Moreover, results are indistinguishable from the averaged MD trajectories.

\subsection{Cycle space}

\begin{figure}[b!]
 \centering
  \includegraphics[width=\linewidth]{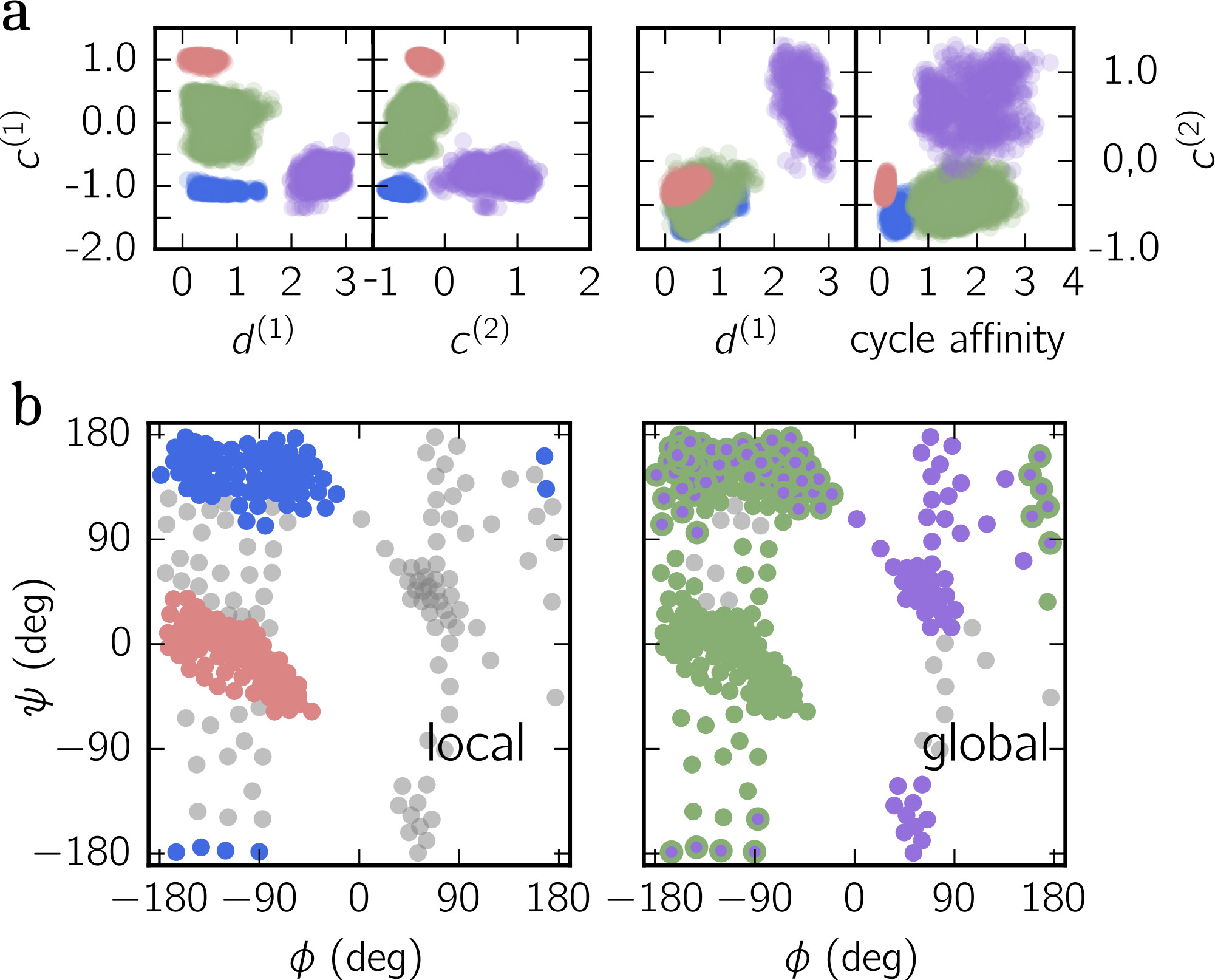}
  \caption{\textbf{Cycle space projections.} (a)~Different cycle space projections are shown for $T=5~$ps. From left to right: The first cycle center dimension versus the first cycle diameter and the second cycle center dimension, and the second cycle center dimension versus the first cycle diameter and the cycle affinity. Colors indicate detected cycle communities. (b)~Centroids (shown in gray) of discretized dihedral space ($\phi,\psi$). Colored circles illustrate centroids that belong to a specific cycle community, as shown in (a). Centroids exhibiting two different colors (edge and face color) are part of two cycle communities. The left panel highlights all centroids belonging to local cycle communities, while the right panel highlights centroids belonging to global communities.}
  \label{fig:cycle_space}
\end{figure}

To infer information about the non-equilibrium dynamics from the cycles detected by the cycle-flux decomposition, we compute the \emph{cycle space} introduced in Sec.~\ref{sec:cycle}, which allows us to identify patterns in the distribution of cycles. To this end, we determine for every identified cycle its center ($c^{(1)}_\alpha,c^{(2)}_\alpha$) and diameter ($d^{(1)}_\alpha, d^{(2)}_\alpha$) in configuration space, and the cycle affinity $A_\alpha$.

In Fig.~\ref{fig:cycle_space}(a), we illustrate four selected projections of these five cycle space coordinates determined for $T=5~$ps, which clearly show that the distribution of cycles indeed exhibits an internal structure and is not random. While the left panels exhibit four distinguishable clusters, the right panels only offer two or three separated point clouds. Our next task is to group cycles (points in cycle space) together forming \emph{cycle communities}. To this end we employ $c$-means clustering (cf. Sec.~\ref{sec:cycle}). For the shown example we find $k=5$ communities to fit the data best (highlighted by different colors).

To understand their significance with respect to the change in molecular configurations, we highlight in dihedral space ($\phi,\psi$) all states that belong to cycles of a specific cycle community by the same color [see Fig.~\ref{fig:cycle_space}(b)]. Apparently, cycles and thus states belonging to the red and blue community coincide with the metastable set $C7_\text{eq}$ and $\alpha_L$, respectively. The purple and green community, on the other hand, connect each two metastable sets, with green cycles linking $C7_\text{eq}$ and $\alpha_L$ configurations and purple cycles linking $\alpha_L$ and $\alpha_R$ configurations. Since the red and blue cycle community only contain states belonging to a single metastable set, we refer to them as \emph{local} cycle communities. On the contrary, cycles that connect two or more metastable sets, here green and purple, are referred to as \emph{global} cycle communities.

\subsection{Coarse-grained model}

At this stage, we understand that cycle communities emerge through the interplay between the underlying potential energy landscape and the oscillating electric field that drives the system out of equilibrium. To be able to effectively examine the non-equilibrium dynamics when changing the oscillation period, we coarse-grain the effective rate matrix $\tW$, returning the coarse-grained rate matrix $\mat W^\text{cg}$ with only a handful of states that represent the dominant (cyclic) dynamics. The basic idea is that all cycles forming a specific community are lumped together into a single cycle, the community representative, and all remaining cycles are deleted. The coarse graining is completed when all nonvanishing transition rates are dynamically consistently rescaled while preserving dominant time scales of the system.

\begin{figure}[t]
  \centering
  \includegraphics[width=\linewidth]{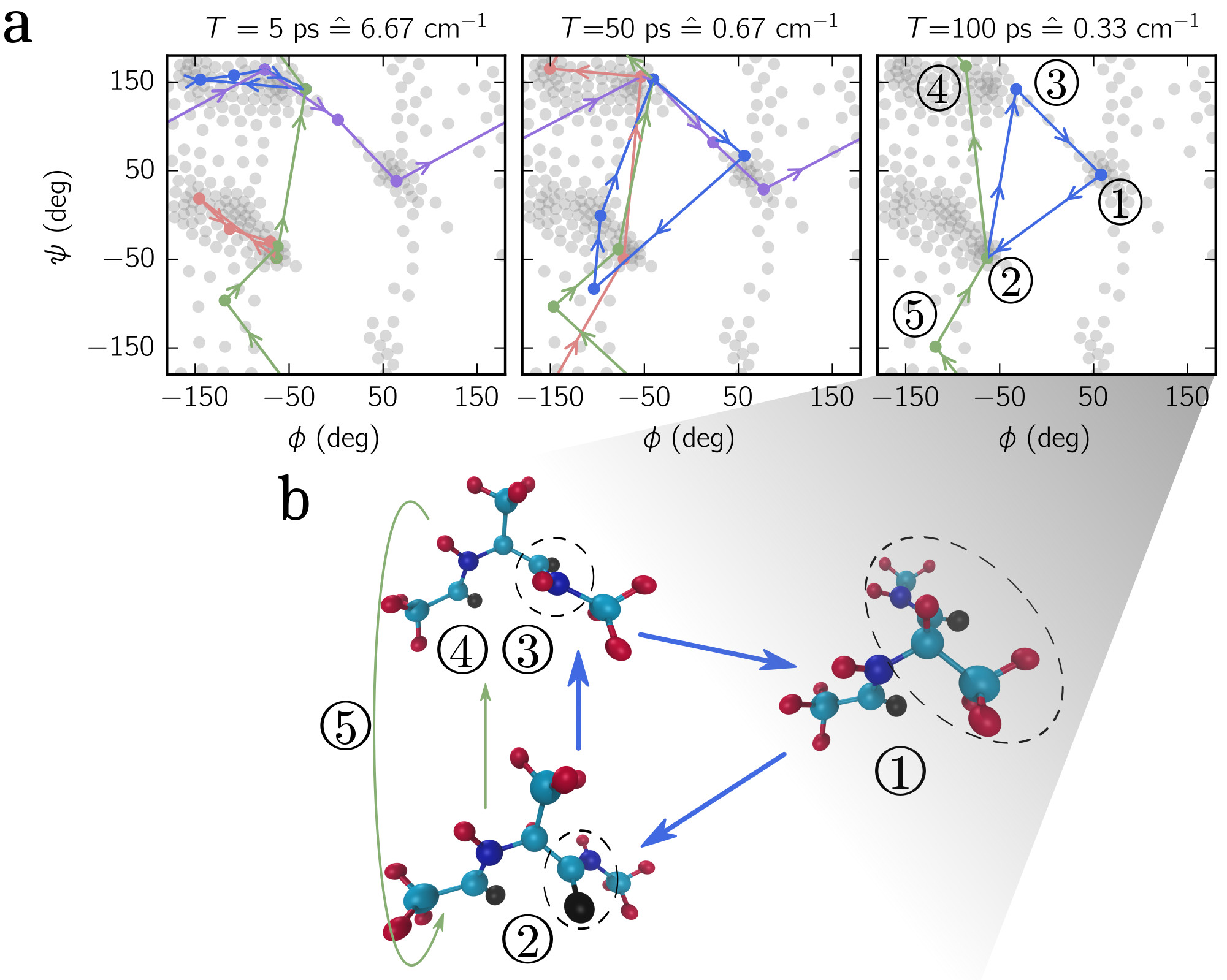}
  \caption{\textbf{Cycle representatives.} (a) Final coarse-grained NE-MSM for different periods $T$ of the applied electric field (also given is the wavenumber). Colors indicate the community representatives. (b) Illustration of the two remaining cycles for driving period $T=100~$ps showing typical configurations. Dashed circles highlight the involved conformational changes.}
  \label{fig:cycle_reps}
 \end{figure}
 
To examine how the dynamics is altered when changing the driving period $T$, we construct coarse-grained NE-MSMs for three different periods: $T = 5~$ps, $50~$ps and $100~$ps. In Fig.~\ref{fig:cycle_reps}(a), the final coarse-grained rate matrices are visualized where colored cycles are representatives of their respective community. For $T=5~$ps the coarse-grained model quantitatively confirms the cycle distribution shown in Fig.~\ref{fig:cycle_space}(b). Interestingly, when the oscillation period is changed, the communities and therefore their representatives change too. For example, for $T=50~$ps the former ($T=5$~ps) local blue and red community vanish and two new communities, also labeled in red and blue, appear. The blue community is of importance as it encloses all three metastable sets $\alpha_R\to\alpha_L\to C7_\text{eq}\to\alpha_R$. For $T=100~$ps no new communities emerge. However, the former ($T=5$~ps) purple and red community disappear, leaving two effective cyclic modes. The green cycle represents a clockwise rotation of the $\psi$ angle, whereas the blue cycle connects all three metastable sets in clockwise direction. Both cycles are illustrated in Fig.~\ref{fig:cycle_reps}(b).


\section{Conclusions}

To conclude, we have introduced a computational method to construct coarse-grained Markov state models with a few states for systems that are periodically driven. This is achieved through exploiting Floquet's theorem and mapping the time-periodic dynamics onto a stationary effective dynamics with the same occupation probabilities. For such a non-equilibrium steady state, we have previously developed a method that is based on coarse-graining in cycle space through clustering, \emph{i.e.}, the identification of similar clusters as measured through suitable collective variables. This approach has the advantage that the entropy production of the effective dynamics is preserved in the coarse-grained model. Through basing the selection of cycle representatives on dynamical data (in particular mean-first passage times) also the dynamics of the coarse-grained system is preserved, a feature that is very hard to achieve in \emph{structural} coarse-graining (\emph{i.e., the reduction of continuous degrees of freedom through combining atoms into beads)}~\cite{mull02}.

As a specific system, we have studied alanine-dipeptide, a paradigm for computational studies of protein dynamics. In a time-dependent electric field, the dynamics averaged over one period breaks detailed balance and becomes cyclic, \emph{i.e.}, transitions have a preferred direction with non-zero probability currents. These cycles can be classified into local cycles residing in a metastable basin, and global cycles connecting these basins. For longer periods, the peptide effectively equilibrates and consequently the local cycles vanish. Our method is not restricted to biomolecules, but could be applied to the stochastic dynamics in other networks.

The major challenge in constructing Markov state models is to identify a suitable low-dimensional space of collective variables (like the cycle center and diameters employed in this work). Major efforts in this field are currently directed towards this dimensionality reduction. In particular machine learning approaches seem to be a promising route to reduce the complexity involved in determining suitable collective variables~\cite{mardt18}. However, such an approach for driven systems will again have to respect the underlying cycle structure, and it will be interesting to see how our results can be incorporated into learning strategies.

\begin{acknowledgments}
  We acknowledge financial support by the Deutsche Forschungsgemeinschaft (DFG) through the collaborative research center TRR 146 (project A7).
\end{acknowledgments}


\appendix

\section{Cycle-flux decomposition}
\label{sec:decomp}

To decompose the current matrix $\mat{J}$, we introduce an algorithm that can be split into two parts. First, we identify nontrivial cycles, \emph{i.e.}, cycles with more than two different states, and secondly we determine their cycle weights. Theoretically, the number of possible cycles in a graph grows exponentially with the number of vertices. However, if both algorithmic steps -- searching for a nontrivial cycle and determine its cycle weight -- are combined by running them alternately, the decomposition becomes computationally affordable even for a large number of states.

To detect a nontrivial cycle, we perform the following steps:
\begin{enumerate}
\item Find the oriented edge $i\to j$ of the largest element of $\mat{J}$: $\operatorname {arg\,max}\,(J_{ij})$.
\item Identify the shortest path (smallest number of transitions) from state $j$ leading back to state $i$ by only following the nonzero transitions of $\mat{J}$. This step can be efficiently realized by applying a breadth-first search~\cite{newman2010}.
\item Return the cycle, $\alpha = \{i\to j \to\, $found path$\}$.
\end{enumerate}
To determine the corresponding cycle weight $\vhi_\alpha$, we consider all fluxes along cycle $\al$ and determine their smallest value, 
\begin{equation}
  \varphi_\alpha \equiv \min\limits_{(i\to j)\ni \alpha} \{J_{ij}\}.
\end{equation}
which becomes the cycle weight $\vhi_\al$.

Summing up both steps, the final cycle-flux decomposition algorithm reads
\begin{enumerate}
\item Find a nontrivial cycle in $\mat{J}$.
\item Compute its cycle weight $\varphi_\alpha$.
\item Update all $J_{ij}$ by subtracting $\varphi_\alpha$ along cycle $\alpha$,  $J_{ij}\leftarrow J_{ij} -\varphi_\alpha,~\forall (i\to j) \in \alpha $ and continue with step 2.
\item The algorithm stops when the residuum $||\mat{J}_{\text{max}}||$ has become smaller than a threshold.
\end{enumerate}
An implementation of this algorithm is provided with the Supplementary Information of Ref.~\cite{knoch2017}.

\section{Cycle representatives}
\label{sec:repr}

All input quantities (\emph{i.e.}, the cycle features affinity, diameter, etc.) are normalized by their variance to make them comparable. These features are then used as input for the fuzzy $c$-means clustering algorithm returning membership degrees $u_{\al l}$ that express the probability that observation $\al$ belongs to community $l$. To obtain an indicator of how good the clustering results are we compute the fuzzy partition coefficient (FPC) that is defined as the Frobenius norm of the membership matrix 
\begin{equation}
  \text{FPC} = \frac{1}{n}\sum_{l=1}^k\sum_{\al=1}^n u_{\al l}^2.
\end{equation}
Here, $k$ is the number of chosen communities and $n$ the number of observations (\emph{i.e.}, number of cycles in our case). The best fitting number of cycle communities is chosen such that the FPC becomes maximal/large. Note that it is not necessarily the best choice to always pick $k$ corresponding to the maximal FPC. Sometimes the determined $k$-value is too low, causing a too coarse description of the cyclic structure. In such a case it is worth checking the second/third best choice.

Once we have determined the number of cycle communities, we require a suitable set of representatives, each representing one community. In order to preserve the large-scale dynamics, we propose the following stochastic algorithm: The general idea is that any set of cycle representatives is thought to be appropriate if the graph spanned by $\mat W^\text{cg}$ is ergodic and the mean first passage times (MFPT) between metastable configurations are preserved. Metastable configurations, if not known beforehand, can be determined for instance by considering all configurations / states found to belong to ``local'' cycle communities, \emph{i.e.}, communities with cycles which states do not overlap with states of other communities [see for example the red and blue colored states in Fig.~\ref{fig:cycle_space}(b)]. Another approach for NE-MSMs is to employ a kinetic clustering scheme using hitting times~\cite{sarich2014,wang2015}, which is in analogy to the often used PCCA+ algorithm~\cite{roblitz2013} but also valid for transition matrices breaking detailed balance. For alanine dipeptide, however, we keep the manual selected metastable configurations as shown in Fig.~\ref{fig:alanine}(c).

Summing up both parts, the full coarse-graining algorithm is outlined as follows:
\begin{enumerate}
\item Pick one representative per cycle community by drawing a random number.
\item Check if the set of cycle representatives span an ergodic (connected) transition network. If yes, determine the coarse-grained transition rates (cf. Sec.~\ref{sec:cg}), else go back to step (1).
\item Compute MFPTs of the coarse-grained model and compare to MFPTs of the full model. If
  \begin{equation}
    \nonumber
    \Bigg|\frac{\text{MFPTs}^\text{cg} -\text{MFPTs}^\text{full}}{\text{MFPTs}^\text{full}} \Bigg| \le \text{threshold},
  \end{equation}
  return $\mat W^\text{cg}$, else go back to step (1).
\end{enumerate}


%

\end{document}